\PassOptionsToPackage{unicode}{hyperref}
\PassOptionsToPackage{hyphens}{url}
\PassOptionsToPackage{dvipsnames,svgnames,x11names}{xcolor}
\documentclass[
]{article}
\usepackage{amsmath,amssymb}
\usepackage[margin=1in]{geometry}
\usepackage{lmodern}
\usepackage{iftex}
\ifPDFTeX
  \usepackage[T1]{fontenc}
  \usepackage[utf8]{inputenc}
  \usepackage{textcomp} % provide euro and other symbols
\else % if luatex or xetex
  \usepackage{unicode-math}
  \defaultfontfeatures{Scale=MatchLowercase}
  \defaultfontfeatures[\rmfamily]{Ligatures=TeX,Scale=1}
\fi
\IfFileExists{upquote.sty}{\usepackage{upquote}}{}
\IfFileExists{microtype.sty}{% use microtype if available
  \usepackage[]{microtype}
  \UseMicrotypeSet[protrusion]{basicmath} % disable protrusion for tt fonts
}{}
\makeatletter
\@ifundefined{KOMAClassName}{% if non-KOMA class
  \IfFileExists{parskip.sty}{%
    \usepackage{parskip}
  }{% else
    \setlength{\parindent}{0pt}
    \setlength{\parskip}{6pt plus 2pt minus 1pt}}
}{% if KOMA class
  \KOMAoptions{parskip=half}}
\makeatother
\usepackage{xcolor}
\newlength{\cslhangindent}
\newlength{\csllabelwidth}
\newlength{\cslentryspacingunit} % times entry-spacing
\newenvironment{CSLReferences}[2] % #1 hanging-ident, #2 entry spacing
 {% don't indent paragraphs
  \setlength{\parindent}{0pt}
  \ifodd #1
  \let\oldpar\par
  \def\par{\hangindent=\cslhangindent\oldpar}
  \fi
  \setlength{\parskip}{#2\cslentryspacingunit}
 }%
 {}
\usepackage{calc}

\ifLuaTeX
\usepackage[bidi=basic]{babel}
\else
\usepackage[bidi=default]{babel}
\fi
\babelprovide[main,import]{american}
\def\languageshorthands#1{}
\ifLuaTeX
  \usepackage{selnolig}  % disable illegal ligatures
\fi
\IfFileExists{bookmark.sty}{\usepackage{bookmark}}{\usepackage{hyperref}}
\IfFileExists{xurl.sty}{\usepackage{xurl}}{} % add URL line breaks if available
\hypersetup{
  pdftitle={cppdlr: Imaginary time calculations using the discrete
Lehmann representation},
  pdfauthor={Jason Kaye, Hugo U. R. Strand, Nils Wentzell},
  pdflang={en-US},
  colorlinks=true,
  linkcolor={Maroon},
  filecolor={Maroon},
  citecolor={Blue},
  urlcolor={Blue},
  pdfcreator={LaTeX via pandoc}}

\title{cppdlr: Imaginary time calculations using \\ the discrete Lehmann
representation}

\usepackage[affil-it]{authblk}
\usepackage{orcidlink}
\author[1,2%
  ]{Jason Kaye%
    \,\orcidlink{0000-0001-8045-6179}\,%
    }
\author[3,4%
  ]{Hugo U. R. Strand%
    \,\orcidlink{0000-0002-7263-4403}\,%
    }
\author[1%
  ]{Nils Wentzell%
    \,\orcidlink{0000-0003-3613-007X}\,%
    }

\affil[1]{Center for Computational Quantum Physics, Flatiron Institute,
New York, NY, USA}
\affil[2]{Center for Computational Mathematics, Flatiron Institute, New
York, NY, USA}
\affil[3]{School of Science and Technology, Örebro University, Örebro,
Sweden}
\affil[4]{Institute for Molecules and Materials, Radboud University,
6525 AJ Nijmegen, the Netherlands}
\date{}

\begin{document}
\maketitle

\hypertarget{summary}{%
\section{Summary}\label{summary}}

Imaginary time Green's functions encode the static and dynamical
response of quantum systems at thermal equilibrium to external
perturbations, such as applied electromagnetic fields. They therefore
represent a direct point of connection between theoretical calculations
and experimental measurements. As a consequence, they appear routinely
in quantum many-body calculations at finite temperature, both for model
systems like the Hubbard model
(\protect\hyperlink{ref-hubbard63}{Hubbard, 1963}), and in ab-initio
electronic structure calculations beyond density functional theory,
e.g., using Hedin's GW method (\protect\hyperlink{ref-golze19}{Golze et
al., 2019}; \protect\hyperlink{ref-hedin65}{Hedin, 1965}). Highly
compact and accurate representations of imaginary time Green's functions
and related imaginary time-dependent response functions are therefore an
important ingredient in the development of robust and efficient codes
for quantum many-body calculations. However, obtaining such
representations has traditionally been challenging, particularly for low
temperature calculations, in which the functions develop steep
gradients.

In the past several years, significant progress has been achieved using
low-rank approximations of the spectral Lehmann representation, which is
given by \[G(\tau) = - \int_{-\infty}^\infty d\omega \,
\frac{e^{-\tau \omega}}{1 + e^{-\beta \omega}} \, \rho(\omega).\] Here,
\(G(\tau)\) is a fermionic single-particle imaginary time Green's
function, and \(\rho(\omega)\) is its corresponding spectral function,
which encodes information about the single-particle excitations of the
underlying quantum many-body system. The spectral function always
exists, but is typically not known. However, the existence of this
integral representation constrains the space of possible imaginary time
Green's functions to lie within the image of the integral operator,
which is numerically low-rank, enabling the construction of highly
compact basis representations. The intermediate representation (IR) was
introduced first, and used the singular value decomposition to obtain an
orthogonal but non-explicit basis of imaginary time Green's functions
(\protect\hyperlink{ref-chikano18}{Chikano et al., 2018};
\protect\hyperlink{ref-shinaoka17}{Shinaoka et al., 2017}). The
recently-introduced discrete Lehmann representation (DLR) uses the
interpolative decomposition to obtain a non-orthogonal basis consisting
of known exponential functions (\protect\hyperlink{ref-kaye22_dlr}{Kaye,
Chen, \& Parcollet, 2022}). The number of basis functions required in
both representations is similar, and typically significantly less than
the previous state-of-the art methods based on orthogonal polynomials
(\protect\hyperlink{ref-boehnke11}{Boehnke et al., 2011};
\protect\hyperlink{ref-dong20}{Dong et al., 2020};
\protect\hyperlink{ref-gull18}{Gull et al., 2018}).

The DLR's use of an explicit basis of simple functions makes many common
operations, including interpolation, integration, Fourier transform, and
convolution, simple and highly efficient. This has led to a variety of
recent algorithmic advances, including in reducing the size of the
Matsubara frequency mesh in dynamical mean-field theory calculations
(\protect\hyperlink{ref-sheng23}{Sheng et al., 2023}), stabilizing the
calculation of the single-particle self-energy via the Dyson equation
(\protect\hyperlink{ref-labollita23}{LaBollita et al., 2023}), improving
the efficiency of the imaginary time discretization in the mixing
Green's function of the Keldysh formalism
(\protect\hyperlink{ref-blommel24}{Blommel et al., 2024};
\protect\hyperlink{ref-kaye23_eqdyson}{Kaye \& Strand, 2023}), and
accelerating the evaluation of imaginary time Feynman diagrams
(\protect\hyperlink{ref-kaye23_diagrams}{Kaye, Strand, et al., 2023}).
It has also yielded immediate applications in computational physics, for
example in low-temperature studies of superconductivity
(\protect\hyperlink{ref-cai22}{Cai et al., 2022};
\protect\hyperlink{ref-hou24}{Hou et al., 2024};
\protect\hyperlink{ref-tanjaroonly23}{Tanjaroon Ly et al., 2023}). The
DLR can be straightforwardly integrated into existing algorithms and
codes, often yielding significant improvements in efficiency, accuracy,
and algorithmic simplicity.

\hypertarget{statement-of-need}{%
\section{Statement of need}\label{statement-of-need}}

\texttt{cppdlr} is a C++ library which constructs the DLR and implements
its standard operations. The flexible yet high-level interface of
\texttt{cppdlr} makes it appealing for use both in small-scale
applications and in existing large-scale software projects. The DLR has
previously been implemented in other programming languages, specifically
in Python via \texttt{pydlr}, in Fortran via \texttt{libdlr}, and in
Julia via \texttt{Lehmann.jl} (\protect\hyperlink{ref-Lehmann.jl}{Chen,
2021}; \protect\hyperlink{ref-kaye22_libdlr}{Kaye, Chen, \& Strand,
2022}; \protect\hyperlink{ref-libdlr}{Kaye \& Strand, 2021};
\protect\hyperlink{ref-pydlr}{Strand \& Kaye, 2021}), as well as in the
\texttt{sparse-ir} library implementing the IR
(\protect\hyperlink{ref-wallerberger23}{Wallerberger et al., 2023}).
\texttt{cppdlr} nevertheless provides a needed platform for future
developments. First, \texttt{cppdlr} is written in C++, a common
language used by many large projects in the quantum many-body physics
community. Second, it offers a high-level user interface simpler than
that of \texttt{libdlr}, enabled by the use of C++ templating and the
\texttt{nda} library (\protect\hyperlink{ref-nda}{{``{nda},''} n.d.})
for array types and BLAS/LAPACK compatibility. These features have, for
example, enabled the implementation of the DLR in the TRIQS library
(\protect\hyperlink{ref-parcollet15}{Parcollet et al., 2015}) for
quantum many-body calculations.

\texttt{cppdlr} is distributed under the Apache License Version 2.0
through a public Git repository
(\protect\hyperlink{ref-cppdlr_git}{Kaye et al., 2023a}). The
project documentation (\protect\hyperlink{ref-cppdlr_doc}{Kaye et al.,
2023b}) is extensive, containing background on the
DLR, a user guide describing example programs packaged with the library,
and application interface (API) reference documentation for all classes
and functions. We envision \texttt{cppdlr} as a platform for future
algorithmic developments involving the DLR, and as a go-to tool for
applications employing the DLR.

\hypertarget{acknowledgements}{%
\section{Acknowledgements}\label{acknowledgements}}

We are thankful for helpful discussions with Kun Chen, Olivier
Parcollet, Malte Rösner, and Yann in 't Veld. H.U.R.S. acknowledges
funding from the European Research Council (ERC) under the European
Union's Horizon 2020 research and innovation programme (Grant agreement
No.~854843-FASTCORR). The Flatiron Institute is a division of the Simons
Foundation.

\bigskip

\hypertarget{refs}{}
\begin{CSLReferences}{1}{0}
\leavevmode\vadjust pre{\hypertarget{ref-blommel24}{}}%
Blommel, T., Kaye, J., Murakami, Y., Gull, E., \& Golež, D. (2024).
\emph{Chirped amplitude mode in photo-excited superconductors}.
\url{https://arxiv.org/abs/2403.01589}

\leavevmode\vadjust pre{\hypertarget{ref-boehnke11}{}}%
Boehnke, L., Hafermann, H., Ferrero, M., Lechermann, F., \& Parcollet,
O. (2011). Orthogonal polynomial representation of imaginary-time
{G}reen's functions. \emph{Phys. Rev. B}, \emph{84}, 075145.
\url{https://doi.org/10.1103/PhysRevB.84.075145}

\leavevmode\vadjust pre{\hypertarget{ref-cai22}{}}%
Cai, X., Wang, T., Prokof'ev, N. V., Svistunov, B. V., \& Chen, K.
(2022). Superconductivity in the uniform electron gas: Irrelevance of
the {K}ohn-{L}uttinger mechanism. \emph{Phys. Rev. B}, \emph{106},
L220502. \url{https://doi.org/10.1103/PhysRevB.106.L220502}

\leavevmode\vadjust pre{\hypertarget{ref-Lehmann.jl}{}}%
Chen, K. (2021). Julia implementation of the discrete {L}ehmann
representation (DLR). In \emph{GitHub repository}. GitHub.
\url{https://github.com/numericaleft/Lehmann.jl}

\leavevmode\vadjust pre{\hypertarget{ref-chikano18}{}}%
Chikano, N., Otsuki, J., \& Shinaoka, H. (2018). Performance analysis of
a physically constructed orthogonal representation of imaginary-time
{G}reen's function. \emph{Phys. Rev. B}, \emph{98}, 035104.
\url{https://doi.org/10.1103/PhysRevB.98.035104}

\leavevmode\vadjust pre{\hypertarget{ref-dong20}{}}%
Dong, X., Zgid, D., Gull, E., \& Strand, H. U. R. (2020).
Legendre-spectral {D}yson equation solver with super-exponential
convergence. \emph{J. Chem. Phys.}, \emph{152}(13), 134107.
\url{https://doi.org/10.1063/5.0003145}

\leavevmode\vadjust pre{\hypertarget{ref-golze19}{}}%
Golze, D., Dvorak, M., \& Rinke, P. (2019). {The \(GW\) Compendium: A
Practical Guide to Theoretical Photoemission Spectroscopy}. \emph{Front.
Chem.}, \emph{7}, 377. \url{https://doi.org/10.3389/fchem.2019.00377}

\leavevmode\vadjust pre{\hypertarget{ref-gull18}{}}%
Gull, E., Iskakov, S., Krivenko, I., Rusakov, A. A., \& Zgid, D. (2018).
Chebyshev polynomial representation of imaginary-time response
functions. \emph{Phys. Rev. B}, \emph{98}, 075127.
\url{https://doi.org/10.1103/PhysRevB.98.075127}

\leavevmode\vadjust pre{\hypertarget{ref-hedin65}{}}%
Hedin, L. (1965). {New Method for Calculating the One-Particle Green's
Function with Application to the Electron-Gas Problem}. \emph{Phys.
Rev.}, \emph{139}, A796--A823.
\url{https://doi.org/10.1103/PhysRev.139.A796}

\leavevmode\vadjust pre{\hypertarget{ref-hou24}{}}%
Hou, P., Cai, X., Wang, T., Deng, Y., Prokof'ev, N. V., Svistunov, B.
V., \& Chen, K. (2024). Precursory {C}ooper flow in ultralow-temperature
superconductors. \emph{Phys. Rev. Res.}, \emph{6}, 013099.
\url{https://doi.org/10.1103/PhysRevResearch.6.013099}

\leavevmode\vadjust pre{\hypertarget{ref-hubbard63}{}}%
Hubbard, J. (1963). Electron correlations in narrow energy bands.
\emph{Proc. R. Soc. Lon. Ser.-A}, \emph{276}(1365), 238--257.

\leavevmode\vadjust pre{\hypertarget{ref-kaye22_dlr}{}}%
Kaye, J., Chen, K., \& Parcollet, O. (2022). Discrete {L}ehmann
representation of imaginary time {G}reen's functions. \emph{Phys. Rev.
B}, \emph{105}, 235115.
\url{https://doi.org/10.1103/PhysRevB.105.235115}

\leavevmode\vadjust pre{\hypertarget{ref-kaye22_libdlr}{}}%
Kaye, J., Chen, K., \& Strand, H. U. R. (2022). {libdlr}: {E}fficient
imaginary time calculations using the discrete {L}ehmann representation.
\emph{Comput. Phys. Commun.}, \emph{280}, 108458.
\url{https://doi.org/10.1016/j.cpc.2022.108458}

\leavevmode\vadjust pre{\hypertarget{ref-libdlr}{}}%
Kaye, J., \& Strand, H. U. R. (2021). Libdr: Imaginary time calculations
using the discrete {L}ehmann representation ({DLR}). In \emph{GitHub
repository}. GitHub. \url{https://github.com/jasonkaye/libdlr}

\leavevmode\vadjust pre{\hypertarget{ref-kaye23_eqdyson}{}}%
Kaye, J., \& Strand, H. U. R. (2023). A fast time domain solver for the
equilibrium {D}yson equation. \emph{Adv. Comput. Math.}, \emph{49}(4).
\url{https://doi.org/10.1007/s10444-023-10067-7}

\leavevmode\vadjust pre{\hypertarget{ref-kaye23_diagrams}{}}%
Kaye, J., Strand, H. U. R., \& Golež, D. (2023). \emph{Decomposing
imaginary time Feynman diagrams using separable basis functions:
Anderson impurity model strong coupling expansion}.
\url{https://arxiv.org/abs/2307.08566}

\leavevmode\vadjust pre{\hypertarget{ref-cppdlr_git}{}}%
Kaye, J., Strand, H. U. R., \& Wentzell, N. (2023a). Cppdlr: Imaginary
time calculations using the discrete {L}ehmann representation. In
\emph{GitHub repository}. GitHub.
\url{https://github.com/flatironinstitute/cppdlr}

\leavevmode\vadjust pre{\hypertarget{ref-cppdlr_doc}{}}%
Kaye, J., Strand, H. U. R., \& Wentzell, N. (2023b). Cppdlr: Imaginary
time calculations using the discrete {L}ehmann representation. In
\emph{GitHub-hosted documentation}.
\url{https://flatironinstitute.github.io/cppdlr/}

\leavevmode\vadjust pre{\hypertarget{ref-labollita23}{}}%
LaBollita, H., Kaye, J., \& Hampel, A. (2023). \emph{Stabilizing the
calculation of the self-energy in dynamical mean-field theory using
constrained residual minimization}.
\url{https://arxiv.org/abs/2310.01266}

\leavevmode\vadjust pre{\hypertarget{ref-nda}{}}%
{nda}: C++ library for multi-dimensional arrays. (n.d.). In \emph{GitHub
repository}. \url{https://github.com/TRIQS/nda}

\leavevmode\vadjust pre{\hypertarget{ref-parcollet15}{}}%
Parcollet, O., Ferrero, M., Ayral, T., Hafermann, H., Krivenko, I.,
Messio, L., \& Seth, P. (2015). {TRIQS}: {A} toolbox for research on
interacting quantum systems. \emph{Comput. Phys. Commun.}, \emph{196},
398--415. \url{https://doi.org/10.1016/j.cpc.2015.04.023}

\leavevmode\vadjust pre{\hypertarget{ref-sheng23}{}}%
Sheng, N., Hampel, A., Beck, S., Parcollet, O., Wentzell, N., Kaye, J.,
\& Chen, K. (2023). Low-rank {G}reen's function representations applied
to dynamical mean-field theory. \emph{Phys. Rev. B}, \emph{107}, 245123.
\url{https://doi.org/10.1103/PhysRevB.107.245123}

\leavevmode\vadjust pre{\hypertarget{ref-shinaoka17}{}}%
Shinaoka, H., Otsuki, J., Ohzeki, M., \& Yoshimi, K. (2017). Compressing
{G}reen's function using intermediate representation between
imaginary-time and real-frequency domains. \emph{Phys. Rev. B},
\emph{96}, 035147. \url{https://doi.org/10.1103/PhysRevB.96.035147}

\leavevmode\vadjust pre{\hypertarget{ref-pydlr}{}}%
Strand, H. U. R., \& Kaye, J. (2021). Pydlr: Imaginary time calculations
using the discrete {L}ehmann representation (DLR). In \emph{Python
Package Index (PyPI) project}. \url{https://pypi.org/project/pydlr/}

\leavevmode\vadjust pre{\hypertarget{ref-tanjaroonly23}{}}%
Tanjaroon Ly, A., Cohen-Stead, B., Malkaruge Costa, S., \& Johnston, S.
(2023). Comparative study of the superconductivity in the {H}olstein and
optical {Su-Schrieffer-Heeger} models. \emph{Phys. Rev. B}, \emph{108},
184501. \url{https://doi.org/10.1103/PhysRevB.108.184501}

\leavevmode\vadjust pre{\hypertarget{ref-wallerberger23}{}}%
Wallerberger, M., Badr, S., Hoshino, S., Huber, S., Kakizawa, F.,
Koretsune, T., Nagai, Y., Nogaki, K., Nomoto, T., Mori, H., Otsuki, J.,
Ozaki, S., Plaikner, T., Sakurai, R., Vogel, C., Witt, N., Yoshimi, K.,
\& Shinaoka, H. (2023). Sparse-ir: Optimal compression and sparse
sampling of many-body propagators. \emph{SoftwareX}, \emph{21}, 101266.
\url{https://doi.org/10.1016/j.softx.2022.101266}

\end{CSLReferences}

\end{document}